\documentstyle[A4,epsf,twocolumn]{article}
\begin{document}

\textwidth18.5cm
\textheight24.5cm
\topmargin-1.5cm
\oddsidemargin-1.0cm

\title{ Testing of quantum phase in matter wave optics}

\author{Jaroslav \v{R}eh\'{a}\v{c}ek$^{\dag}$
, Zden\v{e}k
Hradil$^{\dag}$,
Michael Zawisky$^{\ast}$, Saverio Pascazio$^{\S}$, \\
Helmut Rauch$^{\ast}$,
and Jan Pe\v{r}ina$^{\dag\ddag}$\\~\\
\dag \it Department of Optics, Palack\'{y} University, 17. listopadu 50,
772~07 Olomouc, Czech Republic,\\
$\ast$  \it Atominstitut der \"{O}sterreichischen Universit\"{a}ten,
A-1020 Vienna, Austria,\\
$\S $ \it Dipartimento di Fisica, Universit\`{a} di Bari and
\it Istituto Nazionale di Fisica Nucleare,\\
\it Sezione di Bari, I-70126 Bari,
Italy,\\\it  and \\
$\ddag$ \it Joint Laboratory of Optics, Palack\'{y} University and
Czech Acad. Sci., \\
\it 17. listopadu 50, 772~07 Olomouc, Czech Republic}
\date{}
\maketitle

\begin{abstract}
Various phase concepts may be treated as special cases of
the maximum likelihood estimation.
For example the discrete Fourier estimation
that actually coincides with the operational phase of
Noh, Foug\`{e}res and Mandel is obtained
for continuous Gaussian signals with phase
modulated mean.
Since signals in quantum theory
are discrete, a prediction different from that given by
the Gaussian hypothesis
should be obtained as the best fit assuming a discrete
Poissonian  statistics of the signal.
Although the Gaussian estimation gives a satisfactory
approximation for fitting  the phase distribution of
almost any state the optimal phase estimation offers in certain
cases a measurable better performance. This has been demonstrated
in neutron--optical experiment.
\end{abstract}

\section{Introduction}
Physics enables us to comprehend Nature by considering
intimate  relations between various effects. Any physical
observation can always be compared and analyzed  in relation with a
particular internal model,  providing   us with some additional
insight into the laws of Nature. However, this  effort need
not and usually does not tend to a unique
solution.  It may happen that there are several plausible models
and the given observation is not able to discriminate among them.
On the other hand it may also happen that
some of the assumptions about the system need not to be apparent.
 Certain statements therefore pretend to
be  more general than they are in reality. This interplay between
physics and philosophy may be demonstrated  on the long standing
problem of quantum theory--on the problem of quantum phase.
Phase measurements  belong to standard detection techniques
revealing the wave property of the signal. The quantum phase,
however, encountered  theoretical difficulties  when an adequate
quantum theory has been constructed \cite{royer}.

There are several concepts for the description of phase in quantum
theory at present. For an up to date overview see
\cite{prehled,vlasta}. Some of them are  accenting the theoretical
aspects, other the experimental ones.
The operational approach formulated by Noh, Foug\`{e}res and
Mandel (NFM) \cite{nfm2,nfm}
is motivated by the correspondence principle
in classical wave theory. The interference pattern
is adopted for the scheme where the $\sin$ and $\cos$  function
of the phase shift are measured simultaneously in the
8-port homodyne detection.
An equivalent measurement may be realized on the Mach-Zehnder
interferometer, provided that the measurement of an unknown phase
shift is done with and without an additional $\pi/2$ phase
shifter.
  The NFM scheme  is plausible whenever the
signal behaves like a classical wave since besides the principle
of correspondence, no other assumptions has been used.

 As a particular result, the NFM scheme provides the
optimum result, provided that the statistics of the signals
is represented
by Gaussian statistics with a phase sensitive mean and a phase
insensitive noise. Since realistic signals in the quantum world
do not meet these rather restricting criteria,
phase prediction based on them is not optimum in general.
The difference between Gaussian estimation and optimum
treatments  is caused by
the statistical nature of the phenomena. This can be 
experimentally registered in an interferometer with discrete
Poissonian  signal. Several possible interpretations of
this comparison  are noteworthy:
i) This may be considered as a testing of  the
operational quantum phase prediction. It quantifies how
well the NFM  phase concept fits reality.
ii) It  may be
interpreted as a nontrivial statistically motivated ``quantum
calibration"
of an interferometer.
The visibility of  interference fringes is
usually used for this purpose. However, this criterion
focusses on the wave property of the detected signal only.
The proposed method involves and evaluates the  whole detection
process,
particularly the ability to control phase shift in the
experimental arrangement  and the statistics of the detected signal.
This seems to be in accordance with the pragmatic interpretation
of quantum theory, where the results depend irreducibly  on both
state preparation and measurement.
iii) In the framework of wave--particle duality, the proposed
treatment tries to answer the  question: ``Does the
interfering signal  resemble more discrete  particles  or
classical continuous  waves ?"
iv) It provides an example of indirect observation of several
parameters. Particularly, by detecting the interference fringes,
 the phase shift as well as the visibility may be determined as
fluctuating variables. It provides one of the simplest examples
of the so called ``quantum state reconstruction" procedure.

 This paper is organized as follows. In the second section, the
main idea of comparing the NFM scheme with an optimum phase
prediction is developed.
In the third section, the idea is generalized in order to apply
the scheme for phase measurement in matter wave interferometry.
The fourth section provides the experimental  realization
in neutron interferometry the results obtained.

\section{Statistical formulation of operational phase concepts}

In this section the operational phase concept will be naturally
 embedded into the general scheme of quantum
estimation theory \cite{hel,jones}. A similar approach  has been
already used in   \cite{zdenek-lett,zawisky}. However,  since the
purpose of the detection scheme is
to predict the phase shift after each run,
the  point estimators  of phase corresponding to
 the maximum--likelihood  (MaxLik)
estimation  will be used here \cite{braun,zdenek}.
Assume an ideal device with four output channels
enumerated by indices 3,4,5,6, where the actual values of
intensities are registered in each run. The values fluctuate
 in accordance with  the  statistics of
continuous Gaussian signals.
The mean intensities are  modulated by a phase parameter $\theta$
as
\begin{eqnarray}
\bar I_{3,4} = \frac{I}{2} (1\pm V \cos\theta) \nonumber\\
\bar I_{5,6} = \frac{I}{2} (1\pm V \sin \theta),
\end{eqnarray}
where $I$ and $V$ are total input intensity and
visibility of the interference fringes, respectively.
The energy is splitted symmetrically between all the
output ports. This device represents  nothing else than
a classical wave
picture of the original 8-port homodyne detection scheme.
Equivalently, it also corresponds to
a Mach-Zehnder interferometer, when the measurement
is done for an unknown phase shift together with a zero  and
a $\pi/2$
auxiliary phase shifters.  In this case, the data are not
obtained simultaneously, but should be collected during repeated
experiments.
Provided that a particular combination of outputs $I_3,I_4,I_5,I_6$
has been registered, the phase shift should be inferred.
In accordance with the MaxLik approach \cite{kendall},
the sought-after  phase
shift is given by the  value that maximizes the likelihood
function.
The likelihood  function corresponding to the
detection of given data reads
\begin{eqnarray}
\label{LGauss1}
{\cal  L}(\theta)=\frac{1}{\sigma^4 4\pi^2}
           \exp\left\{\frac{1}{2\sigma^2} \left(
           {-[I_3 - \bar I_3 ]}^2-
           {[ I_4 - \bar I_4 ]}^2- \right. \right.\nonumber\\
         \left. \left.  {-[I_5 - \bar I_5]}^2-
                        {[I_6 - \bar I_6]}^2  \right) \right\} .
\end{eqnarray}
 Here the variation  $\sigma$  represents the phase insensitive
noise of each channel.
The sampling of intensities may  serve for an estimation of
phase shift, the average number of particles and the visibility
simultaneously.
A notation analogous to the definition of phase by
Noh, Foug\`{e}res and Mandel \cite{nfm} can be introduced
as
\begin{eqnarray}
 e^{i \theta_{NFM}}= \frac{I_3-I_4 + i (I_5-I_6)}
{\sqrt{(I_3-I_4)^2 + (I_5-I_6)^2}}
\\
R =  \sqrt{(I_3-I_4)^2 + (I_5-I_6)^2}.
\end{eqnarray}
The likelihood function may be rewritten to the form
\begin{eqnarray}
\label{lik2}
&{\cal L}(\theta,V,I)  \propto&  \exp\bigl\{
- \frac{1}{ 2 \sigma^2} [ I - \frac{1 }{2} \sum_{i=3}^6 I_i]^2
\bigr\} \nonumber  \\
 &\times & \exp\bigl\{- \frac{1}{ 4\sigma^2}
[VI -  R \cos(\theta-\theta_{NFM})]^2 \bigr\}\\
\nonumber &\times&
  \exp\bigl\{
 \frac{1}{ 4\sigma^2}   R^2 \cos^2(\theta-\theta_{NFM}) \bigr\}
\end{eqnarray}
and is  maximized by the choice of  parameters
\begin{eqnarray}  \label{wave}
\theta = \theta_{NFM},
\label{NFM1} \\
V = \min \biggl( \frac{2 R}{\sum_{i=3}^6 I_i}, 1 \biggr),
\label{NFM2}\\
I =  \frac{1 }{2} \sum_{i=3}^6 I_i.
\label{NFM3}
\end{eqnarray}
Hence  the operational phase concept  of Noh, Foug\`{e}res and
Mandel is nothing but the
MaxLik estimation  for waves represented by
continuous  Gaussian signal with phase--independent and
symmetrical noises.
These   rather strong assumptions
are associated with the behaviour of waves
in classical field  theory.

Since realistic signals are discrete they meet neither of these
criteria and therefore, deviations in the optimum phase
prediction should be expected.
%The optimum prediction for realistic estimation will be forecast.
Assume  the Poissonian statistics of an ideal laser.
Together with the phase, all  the parameters which are  not
controlled in the experiment will be optimally predicted as well.
Denote for concreteness the detected discrete values as numbers
$n_3,n_4,n_5,n_6.$
The  likelihood function corresponding to this particular
detection
as a function of the parameters $\theta,  V $ and $ N$ reads
\begin{eqnarray}
\label{lik}
&&{\cal L}(\theta,V,N)\propto
\left(\frac{N}{2}\right)^{ n_3+n_4+n_5+n_6 }
e^{-2N} (1-V \cos\theta)^{n_3}
\nonumber \\&&  \times  (1+V\cos\theta)^{n_4}
(1-V\sin\theta)^{n_5}(1+V\sin\theta)^{n_6} .
\end{eqnarray}
The MaxLik estimation for parameters gives   the optimum values
for  the phase  shift, the visibility and the mean particle number as
\begin{eqnarray}
  \label{particle}
e^{i\theta}=\frac{1}{V}
 \left[
   \frac{n_4-n_3}{n_4+n_3}+i\frac{n_6-n_5}{n_6+n_5}
 \right],
\label{max1} \\
V~=\sqrt{{\left( \frac{n_4-n_3}{n_4+n_3}\right)}^2+
        {\left( \frac{n_6-n_5}{n_6+n_5}\right)}^2 },
\label{max2} \\
N = \frac{n_3+n_4 +n_5 +n_6}{2}.
\label{max3}
\end{eqnarray}
These relations provide a  correction  of the Gaussian wave theory with
respect to the discrete signals.
Besides the phase shift, the visibility  of the
interference fringes and the total energy input
can be evaluated simultaneously.

An apparent difference between relations (\ref{NFM1}--\ref{NFM3})
and (\ref{max1}--\ref{max3})  represents the theoretical
background  of the presented treatment.
Adopting the interpretation  of Ref. \cite{nfm},
in both these approaches the unnormalized
$\cos$  and $\sin$ functions of the
phase shift are detected. However, the normalizations  differ in
both approaches. In the former Gaussian case, the normalization is
performed only once, whereas in the latter  Poissonian case
it is done in two steps.
The $\cos $  and $ \sin$  functions of phase are normalized
separately with respect to the total number of particles on both
the
output ports and then again among themselves.
Obviously, both predictions will coincide provided that there is
almost no information  available in the low field limit $I,N
\rightarrow 0.$
Similarly in the strong field limit $I,N
\rightarrow \infty ,$ where any statistics
approaches the  Gaussian one,  the differences must disappear.
Possible deviations  may appear in the intermediate regime,
characterized approximately by  conditions $I, N \approx O(1).$
The test of the difference  between  $(\ref{NFM1}) $
and $(\ref{max1})$ is proposed  as controlled   phase
measurement.  The phase difference may be adjusted to a certain
value and estimated independently using  both the methods
 $(\ref{NFM1}) $
and $(\ref{max1})$ in repeated experiments.
The efficiency of both methods is then compared by evaluation
of confidence intervals.
Since any imperfections of the detection scheme  will smoothen the
differences, it is questionable whether both schemes can be
experimentally distinguished.  This idea will be pursued in the
following sections.

Before doing this, the statistical analysis may  clarify    some
subtle points of the NFM treatment, particularly then the nature of
discarded
data. Obviously the data
yielding the ambiguous phase  $ e^{i\theta_{NFM}}= 0/0$
in equation (\ref{NFM1})
would provide zero visibility  for both (\ref{NFM2}) and
(\ref{max2}). For a more detailed analysis, the Bayes theorem
 may be applied as well. In this case the  likelihood functions
 quantifies the phase information involved in the detected data
as posterior phase distribution.
 Gaussian statistics   provides homogeneous  posterior phase
distribution, whereas Poissonian statistics yields the four
peak  posterior  distribution of the phase shift resembling effectively  the
homogeneous one.  This statistical   analysis supports the
conclusion of Refs. \cite{comment,Richter,Pegg} that
the detected data cannot be discarded. Provided that some data
are discarded, the average number of particles, i.e. the average
energy corresponding to the phase detection changes.
Particularly, provided that the experiment has been done $M$ times
and a total number $M_i$  of particles has been detected in each run,
the average number of particles is simply $\sum_{i, all} M_i/ M.$
Provided that some data appearing $M_{d}$ times
 are discarded and not included in the
evaluation  the corresponding phase estimation is done with
an average intensity loss  $(\sum_{i, all} M_i - \sum_{i, disc} M_i)/
(M-M_d).$  Obviously, the measurement with and without discarding
is done with a different energy input and this difference may be
substantial. For example, this explains the  ambiguity of the
interpretation in Ref. \cite{TorgMand}. When the phase of a quantum
field is measured against a classical field, no data are
discarded since the field is strong.
The relative phase of two quantum fields may then be
evaluated  separately against a strong field  and a
difference of two such phases values  provides the relative phase.
However, this cannot be directly compared with direct measurements
of two (weak) quantum fields against themselves, when some data are
canceled. Although both measurements have been done with the same
quantum states, the average  energies of the corresponding phase
detection may differ significantly. No wonder that the
measurement with higher average energy  gives better (sharper)
results as observed in Ref. \cite{TorgMand}.

The analysis given here will be adopted to the case of neutron
interferometry, when the number  of particles  can be detected
and discriminated with high efficiency \cite{rauch}.
Highly efficient neutron detectors then provide almost perfect
neutron number measurements.  However the interference pattern
exhibits lower visibility and the splitting process is far away from
the symmetrical case.  This will be taken into account
when adopting the scheme to the case of neutron interferometry.

Analogous analysis for detection of light needs other modification.
Since the efficiency of photodetection is less than unity, number of
particles cannot be detected in such a straightforward way. Instead, the
absence or presence of the week signal on the detector can be used for
phase estimation. However, this will be dealt with separately in the
forthcoming publication.

\section{Phase estimation in neutron interferometry}

Assume the following modification of the scheme developed above.
The output ports of an interferometer are considered to be
nonsymmetrical. The measurement is done with many auxiliary phase
shifters.
Neutron beams inside the perfect crystal neutron
interferometer will be described at first as a  classical wave.
The signals $I_j^o$, $I_j^h$ detected at
the two output ports $o$,$h$ (Fig.~1) are regarded as stochastic Gaussian
intensities with phase sensitive means
%%%%%%%%%%%%%%%%%%%%%%%%%%%%%%%%%%%%%%%%%%%%%%%%%%%%%%%%%%%%%%%%%%%%%%
\begin{eqnarray} \label{stredni}
\bar{I}_j^o&=& I^o+I^V\cos(\theta+\Delta_j) \nonumber \\
\bar{I}_j^h&=& I^h-I^V\cos(\theta+\Delta_j),
\end{eqnarray}
%%%%%%%%%%%%%%%%%%%%%%%%%%%%%%%%%%%%%%%%%%%%%%%%%%%%%%%%%%%%%%%%%%%%%%
where $\theta$ is the true phase shift between the two
branches of the interferometer, $I^o$ and $I^h$ are the mean
intensities of the two interference fringes, $I^V$ stands
for the modulation amplitude of the interference fringes
(unnormalized visibility) and $\Delta_j$ are the
values of the auxiliary shifts
%%%%%%%%%%%%%%%%%%%%%%%%%%%%%%%%%%%%%%%%%%%%%%%%%%%%%%%%%%%%%%%%%%%%%%
\begin{equation} \label{posuvy}
\Delta_j=\frac{2\pi j}{N}, \qquad j=0..N-1, \qquad N\in {\cal N}.
\end{equation}
%%%%%%%%%%%%%%%%%%%%%%%%%%%%%%%%%%%%%%%%%%%%%%%%%%%%%%%%%%%%%%%%%%%%%%%
The Gaussian statistics of the detected signal yields its
likelihood function in the form
%%%%%%%%%%%%%%%%%%%%%%%%%%%%%%%%%%%%%%%%%%%%%%%%%%%%%%%%%%%%%%%%%%%%%%%
\begin{equation} \label{lf}
{\cal L}(\theta)\propto\exp\biggl\{-\frac{1}{2\sigma^2}
\sum\limits_{j=0}^{N-1}\left[(I_j^o-\bar{I}_j^o)^2+
(I_j^h-\bar{I}_j^h)^2\right]\biggr\}.
\end{equation}
%%%%%%%%%%%%%%%%%%%%%%%%%%%%%%%%%%%%%%%%%%%%%%%%%%%%%%%%%%%%%%%%%%%%%%%
Introducing the complex parameter R as
\begin{eqnarray} \label{R}
R&=&  \sum\limits_j(I_j^o-I_j^h)\exp(-i\Delta_j),
\end{eqnarray}
the parameters maximizing the likelihood function read
%%%%%%%%%%%%%%%%%%%%%%%%%%%%%%%%%%%%%%%%%%%%%%%%%%%%%%%%%%%%%%%%%%%%%%%
\begin{eqnarray} \label{thetaNFM}
e^{i\theta}&=&\frac{R}{|R|},
\end{eqnarray}
\begin{equation} \label{Im}
I^o= \frac{1}{N}\sum\limits_j I_j^o, \qquad
I^h= \frac{1}{N}\sum\limits_j I_j^h,
\end{equation}
\begin{eqnarray} \label{V_final}
I^V&=&\mbox{min}\biggl\{ \frac{|R|}{N}, I^o ,I^h  \biggr\}.
\end{eqnarray}
The relation (\ref{V_final})  follows from the condition of
semidefiniteness of the amplitude  $I^V\leq\mbox{min}\{I^o,I^h\}$.
Expression (\ref{thetaNFM}) represents  a generalization
of the NFM formula (\ref{NFM1}), which may be
recovered provided that the measurement is done  at the two
positions $\Delta_0=0$, $\Delta_1=\pi/2$ only.
Sometimes it happens that $R=0$ for recorded data.
In this case the data are phase insensitive  yielding
$ I^V=0 $ and the posterior phase distribution is
homogeneous.

Notice that this approach is  well known in optics  as
  phase estimator  of  discrete Fourier transformation
(DFT) \cite{goodman}.  Define for a  discrete signal $F_j$
its DFT as
%%%%%%%%%%%%%%%%%%%%%%%%%%%%%%%%%%%%%%%%%%%%%%%%%%%%%%%%%%%%%%%%%%%%%%%
\begin{eqnarray} \label{four}
X(m)&=&\frac{1}{N}\sum\limits_{j=0}^{N-1}F_j
\exp(-i2\pi jm/N).
\end{eqnarray}
%%%%%%%%%%%%%%%%%%%%%%%%%%%%%%%%%%%%%%%%%%%%%%%%%%%%%%%%%%%%%%%%%%%%%%
As follows from the comparison of relations
 (\ref{four}), (\ref{thetaNFM}) and
(\ref{V_final}), the
 visibility and phase of the generalized NFM treatment corresponds
 to
the modulus and argument  of the complex coefficient $ X(1)\equiv
R$
%%%%%%%%%%%%%%%%%%%%%%%%%%%%%%%%%%%%%%%%%%%%%%%%%%%%%%%%%%%%%%%%%%%%%%
\begin{equation} \label{znovu}
\theta= \arg\{X(1)\}, \qquad I^V=|X(1)|.
\end{equation}
%%%%%%%%%%%%%%%%%%%%%%%%%%%%%%%%%%%%%%%%%%%%%%%%%%%%%%%%%%%%%%%%%%%%%%
The   discrete signal corresponds to the  difference of
registered discrete scans of interference fringes
created by changing the auxiliary shift
%%%%%%%%%%%%%%%%%%%%%%%%%%%%%%%%%%%%%%%%%%%%%%%%%%%%%%%%%%%%%%%%%%%%%%
\begin{equation} \label{subst}
F_j=I_j^o-I_j^h.
\end{equation}
%%%%%%%%%%%%%%%%%%%%%%%%%%%%%%%%%%%%%%%%%%%%%%%%%%%%%%%%%%%%%%%%%%%%%%
The frequency $m$ of the interference
fringe  in (\ref{znovu}) is one
since the fringe has been scanned only once  (\ref{posuvy}).

The generalization NFM scheme  given by  phase estimation
    (\ref{thetaNFM})
is not optimal provided that the  detected  signal is Poissonian. In
this case the likelihood function reads
%%%%%%%%%%%%%%%%%%%%%%%%%%%%%%%%%%%%%%%%%%%%%%%%%%%%%%%%%%%%%%%%%%
\begin{eqnarray}    \label{Lpoiss}
\cal{L}(\theta)&\propto&\prod\limits_{j=0}^{N-1}
\prod\limits_{\alpha=o,h} \bigl(  \bar{I}_j^{{\alpha}} \bigr)^{n_j^{\alpha}}
e^{-\bar{I}_j^{\alpha}},
\end{eqnarray}
%%%%%%%%%%%%%%%%%%%%%%%%%%%%%%%%%%%%%%%%%%%%%%%%%%%%%%%%%%%%%%%%%%
where the mean output intensities $\bar{I}_j^{\alpha}$ are given
by (\ref{stredni}) and $n_j^{\alpha}$ represent the number of
detected neutrons.
 Unfortunately, the explicit relations  analogous
to  (\ref{max1}--\ref{max3}) cannot be found
analytically. The analysis  must therefore be carried out numerically.
However, various limiting and asymptotic cases can be  discussed
analytically  as shown in \cite{goodman}.
 DFT estimation
is the best estimation available provided a large number of
particles is detected.
No difference between Gaussian and Poissonian cases
 can be therefore expected  in the regime of high intensities
$I_o,I_h\gg 1$.
In the opposite case of low output intensities $I_o,I_h\ll 1$,
the most frequent samples are ``no detection at all'' and
``one neutron detected'' in some beam at some value
of the auxiliary shift. Direct substitution of this
samples in both the Poissonian and the Gaussian likelihood
functions  tends to the prediction with undefined phase
with $I^V=0.$

\begin{figure}
\vspace*{3.5cm}
\hspace*{-2cm}
\epsfxsize=11cm \epsfbox{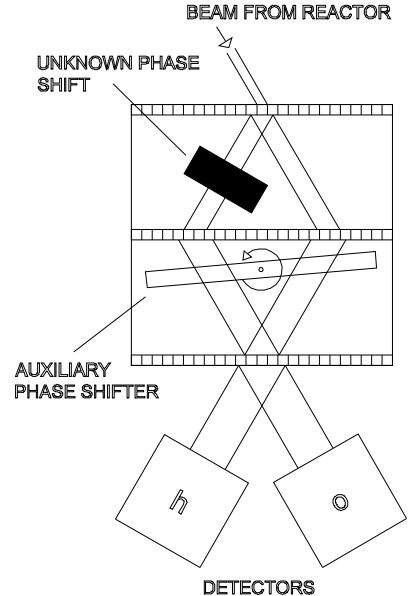}
\caption{Outline of the experimental setup.}
\end{figure}

\begin{figure*}
\vspace*{-3cm}
\epsfysize=25cm \epsfbox{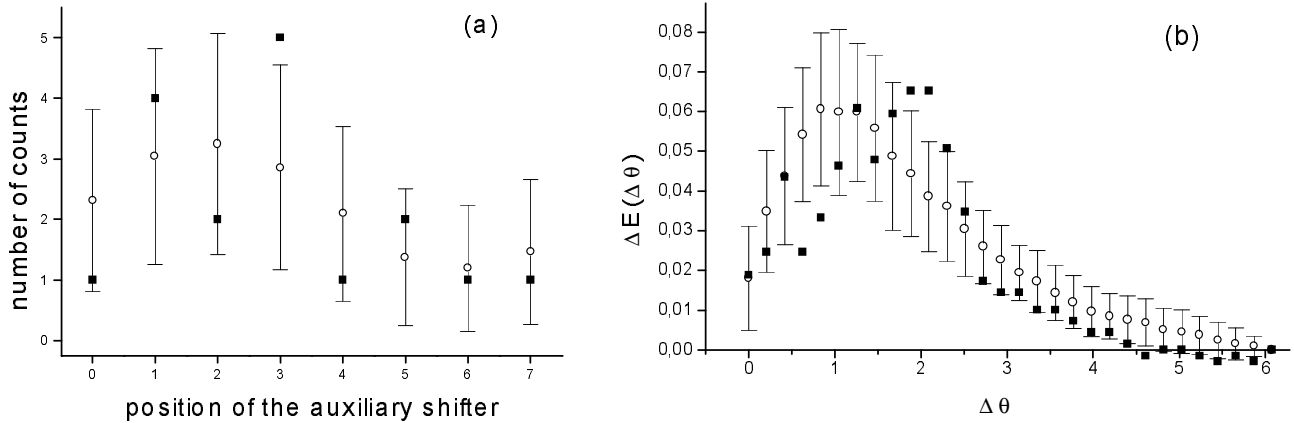}
\vspace*{-16cm}
\caption{(a)Detected interference fringe $\bar{I}^o_j$
as mean of $690$
single measurements ($\circ$) and $68.3\%$ error bars
 for numerical values $I^o=2.21$ neutrons,
$I^h=6.33$ neutrons, $I^V=1.03$ neutrons and $\theta=4.83$ rad.
A typical single detection $I^o_j$ denoted by full squares
is shown as an example.
(b)  Experimentally obtained
$\Delta E(\Delta\theta)$  denoted by full squares
are compared with theoretical prediction
denoted by corresponding mean values
($\circ$) and error bars for $690$ samples.}
\end{figure*}

Similarly to the previous case  of strong field, where  both methods
are equally good,  in the low field  limit both methods are equally bad.
Similarly in the
 limit of low visibility $I^V\ll\mbox{min}\{I_o,I_h\}$,
the results obtained with the use of the true Poissonian statistics
are virtually identical to those yielded by the DFT \cite{goodman}.
No  difference between the wave and
quantum approach can be observed here.
 The difference between the predictions
for the quantum and wave phase estimation in a
realistic experiment should become significant
 for visibility  close to unity
and average number of detected particle of order one.

\section{Experiment}

Our experiments were performed at the neutron interferometry setup
at the 250 kW TRIGA reactor in Vienna.
Thermal neutrons which are emitted from the moderator of the reactor
behave like statistically
independent particles.
Therefore the correct description of the counting
statistics of the input beam and both output beams is a
Poissonian distribution \cite{rauch}.
Fig. 1 shows the experimental arrangement.
The input beam is split by the perfect crystal
interferometer into two partially  coherent beams.
One of the beams passes a phase plate (grey shaded region)
which introduces an unknown phase shift
 which has to be estimated from the experimental data.
Then both beams are passing an auxiliary phase shifter
which modulates
the output intensities at the detectors $o$ and $h$.
At the two output ports BF3 gas detectors enable
single neutron counting with nearly 100 per cent efficiency.
The very low intensities at the outgoing beams
(1 neutron per second) allow a comfortable electronical separation of the
detector pulses. The mean number of collected neutrons is a linear
function of the counting time which enables an
adjustment of the desired intensities by proper
selection of the counting time.
In neutron interferometry an auxiliary phase shifter can
be rotated in several discrete positions denoted by indices $j$
in the intensity equation.
Unique phase estimation is achieved even when other parameters
of the setup (e.g. the mean intensities, the visibility and the
frequency of the oscillation pattern) are unknown.
In our experiment 8 equidistant positions of the phase shifter
were used for  generation of the intensity modulation.

\begin{figure*}
\protect\vspace*{-0.3cm}
\hspace*{2cm}
\epsfysize=19cm \epsfbox{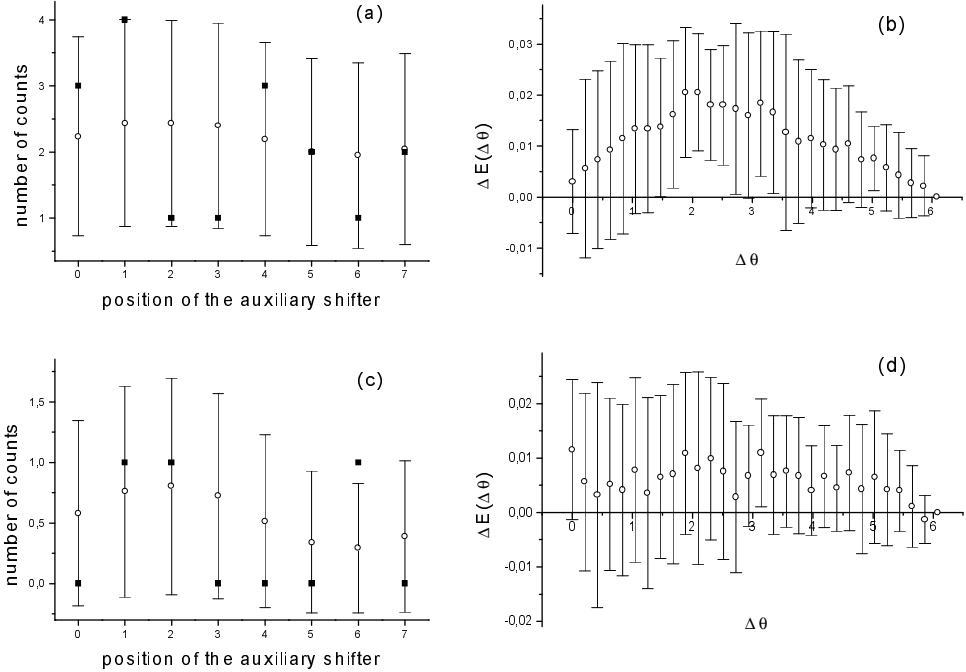}
\protect\vspace*{-9.5cm}
\caption{(a)
Interference fringe $\bar{I}^o_j$ ($\circ$) and error bars
corresponding to computer simulation of the experiment
with numerical values $I^o=2.21$ neutrons, $I^h=6.33$
neutrons, $I^V=0.258$ neutrons and $\theta=4.83$ rad.
A typical sample denoted by full squares
is shown as an example.
(b) $\Delta E(\Delta\theta)$ corresponding to the same computer
simulation  is evaluated. Mean values ($\circ$) and error bars
corresponding to $690$ samples are given.
The interpretation of  panels (c)  and (d) is the same as for
(a) and (b), respectively. Numerical  values are
 $I^o=0.551$ neutrons, $I^h=1.582$ neutrons,
$I^V=0.258$ neutrons and $\theta=4.83$ rad.}
\vspace*{-0.4cm}
\end{figure*}
\nobreak

To compare the efficiency of the NFM phase predictions with the  optimum
Poissonian MaxLik estimation the following procedure has been chosen.
Each sample of data consisting of the number of neutrons
counted in beams $o, h$ in eight positions $\Delta_0, \ldots, \Delta_7$
(\ref{posuvy}), was processed using NFM formula
(\ref{thetaNFM}) resulting in phase prediction $\theta_{N\!F\!M}$.
The relative frequency $f_g(\Delta\theta)$
characterizes  how many times  the  estimated phase
$\theta_{N\!F\!M}$ falls into the
chosen phase window $\Delta\theta$ (confidence interval) around the
true phase shift. The same procedure was repeated for phase
prediction based on numerical maximization
of the Poissonian likelihood function (\ref{Lpoiss})
\cite{zawisky,zdenek-lett} yielding the relative frequency of
``hits'' $f_p(\Delta\theta)$. The quantity
%%%%%%%%%%%%%%%%%%%%%%%%%%%%%%%%%%%%%%%%%%%%%%%%%%%%%%%%%%%%%%%%%%%%
\begin{eqnarray} \label{diff}
\Delta E(\Delta\theta)&=&f_p(\Delta\theta)-f_g(\Delta\theta)
\end{eqnarray}
%%%%%%%%%%%%%%%%%%%%%%%%%%%%%%%%%%%%%%%%%%%%%%%%%%%%%%%%%%%%%%%%%%%%
represents the difference in efficiency of the quantum and wave phase
estimations for the given  phase window  $\Delta\theta$.
If this quantity is found to be positive, it means that the MaxLik
estimation is better than its Gaussian counterpart (simply because more
estimates of the phase shift fall into the chosen angular window if the
former procedure is followed). If, on the other hand, this quantity is
not different from zero (in a statistically significant way) the two
data evaluation procedures are statistically equivalent and no
discrimination is possible.

The result of the data analysis is shown in Figs.~2--4.
Each figure consists of two different parts.
The left panels show the detected (or simulated) data.
The right panels then provide the interpretation of the
corresponding data. Analysis of the experimental
data is summarized in Fig.~2. In addition to experiment, two
Monte--Carlo simulations have been performed simulating experimental
conditions under which the difference between Poissonian and
Gaussian predictions is negligible.
The result of the simulations is shown in Fig.~3.
Finally, the possibility to estimate several parameters simultaneously
is illustrated in Fig.~4.
The difference $\Delta E$ was calculated using $690$
experimental samples measured in experiment with
average beam intensities $I^o$=2.2 neutrons, $I^h$=6.3
neutrons and visibility normalized with respect to $o$-beam being
about $47\%$. As already explained in Section 2, this represents
a critical situation, because even though there are less than $10$
counts in each experimental run, one is nonetheless trying to get an
estimation of the value of the phase shift.
The experimental values of
$\Delta E$ are depicted in Fig.~2b by full
squares. For comparison, a theoretical prediction corresponding
to the same values of parameters as in the real experiment
were simulated in the Monte--Carlo experiment  using  $40000$
samples. Open circles in the Fig.~2b show the
corresponding mean values of the difference.
However, since the  experimental data   are limited due to the
experimental conditions and available  time to the relatively small
number of 690 samples, the real data are fluctuating around \
the mean values.
Statistical significance of the
experimental  results is demonstrated again using Monte--Carlo
simulations.
Other 20 simulations has been done, each of them with 690 samples
The variance of the ensemble $\{\Delta E_j\}, $  is shown in the
Fig.~2b as ``error bars'' for each phase window.

\begin{figure*}
\protect\vspace*{-2.5cm}
\hspace*{0cm}
\epsfysize=25cm \epsfbox{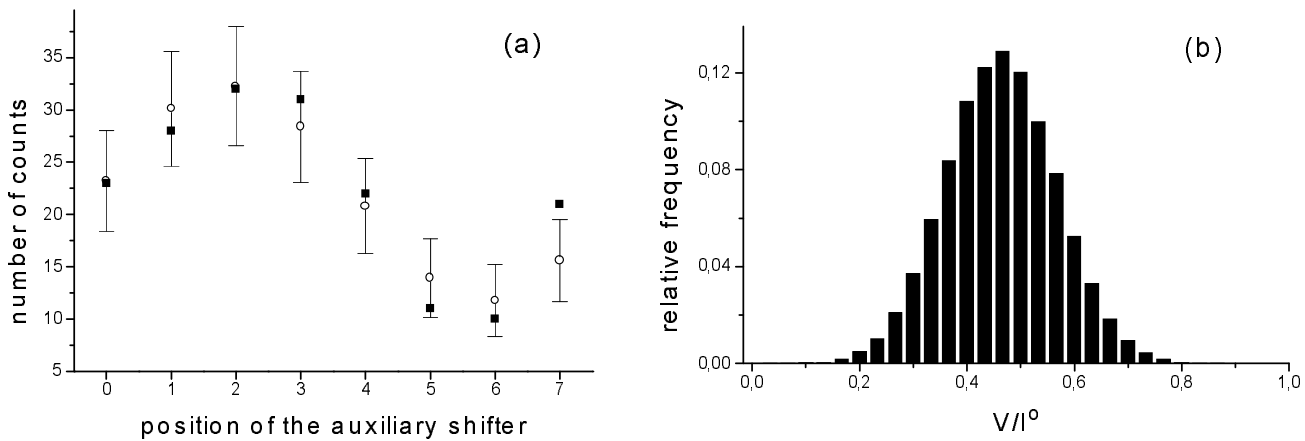}
\protect\vspace*{-16cm}
\caption{(a)
Interference fringe $\bar{I}^o_j$ ($\circ$) and error bars
corresponding to a computer simulation of the experiment
with numerical values $I^o=22.1$ neutrons, $I^h=63.3$ neutrons,
$I^V=10.3$ neutrons and $\theta=4.83$ rad. A typical
sample is denoted by full squares as an example.
(b)
Histogram of estimated visibilities normalized with
respect to $o$-beam  obtained in  the same simulation.
The true normalized visibility is $46.7\%$.}
\vspace*{0cm}
\end{figure*}

A significant difference between the
effectiveness of  classical  and optimum  treatments is apparent in
Fig.~2. The optimum estimation provides an improvement in
fitting of the phase shift and the  difference is beyond the
estimation error, approximately  2.5 standard deviations in the
optimum case.
Obviously, no better performance of the MaxLik method can be expected
for large values of the phase window $\Delta\theta$ (any sensible
statistical method would yield rather reasonable results). Likewise, no
real improvement over the Gaussian estimate can be expected when
$\Delta\theta$ is close to zero, because too few data would then be
accepted. The existence of a ``best choice'' for the phase window is
therefore in itself an interesting feature of the method we propose.
However notice that  generalized NFM scheme
(or equivalently DFT phase estimator) fits the phase shift
quite well. The most pronounced difference is
about $6\%$  in the window of width about $1.256$ rad.
For example it means that the Gaussian
fitting procedure  hits this window $442$ times, whereas the
Poissonian one $484$ times from a total number of $690$ events.
The experimental difference in the  score  is
therefore $42$ i.e. $6\%$ in favour of the latter method.
This difference is a random  number and theory predicts  its
value as $41 \pm 12$. The observed difference is therefore
not large, yet statistically significant.

In most cases, however, the difference between discrete and
continuous nature of the signal is subtle enough to be hidden in
various imperfections of the experimental setup. For
illustration, a computer simulation of two experimental setups
has been performed similarly to the above--mentioned case.
A data analysis of the Monte-Carlo experiment with visibility as
small as $1/4$ of the visibility in Fig.~2 is shown in Fig.~3b.
No statistically significant discrimination between the Gaussian and
Poissonian methods is possible in this case.
The result of a simulation of an experiment with 4-times less
energy of interfering beams is shown  in Fig.~3d.
Also in this case, no discrimination is possible.
In spite of the apparent interference patterns the
classical description of phase can be  fully justified both in
the cases
of low visibility and low intensity. Particularly
the latter case may appear as counterintuitive, since measurement
with small number of particles is traditionally considered as  a
domain of quantum physics!

Unlike the case of phase, the  MaxLik estimation of the visibility
is strongly biased in the case of small intensities.  There is simple
explanation for this behaviour. In low-intensity regime the character
of individual detected samples is determined rather by fluctuations
then by actual parameters of the experimental setup  as seen in
Fig.~3c for example.
The MaxLik estimation  of the visibility fits this fluctuations
and, as a consequence, the estimation is biased.
This is particularly
obvious in the case $I^V= 0$, when interference pattern disappears,
but MaxLik fitting yields a  visibility of about $40\%$ due to
fluctuations.
Nevertheless, except the case of  very low
intensities, the  estimated visibility is meaningful as
demonstrated in Fig.~4, where a histogram of the estimated
visibilities normalized with respect to $o$--beam is shown as
result of a simulation with $10$ times stronger
output beams compared to the output beams used in the experiment.
It is apparent that the estimated visibilities are
distributed with Gaussian--like shape around the value of the
true visibility.

\section{Conclusion}

A statistically motivated analysis of neutron  interferometry
provides a correction to the previously introduced operational quantum
phase concept.  Since the standard approach has been universally
derived, without considering the statistics of the interfering fields,
it cannot be optimal. This additional knowledge may be used for
improved predictions and testing.
This scheme provides therefore a statistically motivated
evaluation of the  whole interferometric system. Instead of the
question of the wave theory: ``How precisely can the interference
 fringes  be distinguished?" a more sophisticated question is
here formulated as: ``What  statistical   properties can be
recognized from an interference pattern?"
In particular, the    experiment performed with neutrons
demonstrated a measurable improvement of phase fitting  for  discrete
Poissonian signals.

\vspace*{10pt}
\noindent
{\large \bf Acknowledgments}

We acknowledge support by the TMR Network ERB FMRXCT 96-0057 "Perfect
Crystal Neutron Optics" of the European Union, by the grant No VS96028 of
Czech Ministry of Education and by the East-West program of the
Austrian Academy of Science.

\end{document}